\documentclass[prd, twocolumn]{revtex4}
\usepackage{dcolumn}
\usepackage{multirow}
\usepackage{graphicx}
\usepackage{amssymb}
\usepackage{bm}
\usepackage{hyperref}
\usepackage{epstopdf}
\usepackage{color}
\usepackage{mathrsfs}
\usepackage{amsmath,amssymb,amsthm}
\begin{document}
\title{Machine Learning Cosmic Expansion History}
\author{Deng Wang}
\email{cstar@mail.nankai.edu.cn}
\affiliation{Theoretical Physics Division, Chern Institute of Mathematics, Nankai University,
Tianjin 300071, China}
\author{Wei Zhang}
\email{cosmoszhang@mail.nankai.edu.cn}
\affiliation{{Department of Physics, Nankai University, Tianjin 300071, China}}
\begin{abstract}
We use the machine learning techniques, for the first time, to study the background evolution of the universe in light of 30 cosmic chronometers. From 7 machine learning algorithms, using the principle of mean squared error minimization on testing set, we find that Bayesian ridge regression is the optimal method to extract the information from cosmic chronometers. By use of  a power-law polynomial expansion, we obtain the first Hubble constant estimation $H_0=65.95^{+6.98}_{-6.36}$ km s$^{-1}$ Mpc$^{-1}$ from machine learning. From the view of machine learning, we may rule out a large number of cosmological models, the number of physical parameters of which containing $H_0$ is larger than 3. Very importantly and interestingly, we find that the parameter spaces of 3 specific cosmological models can all be clearly compressed by considering both their explanation and generalization abilities.

\end{abstract}
\maketitle

\textit{Introduction}.--- During the past two decades, with gradually mounting data, a large number of cosmological observations such as Type Ia supernova (SNIa) \cite{1,2}, cosmic microwave background (CMB) anisotropies \cite{3,4} and baryonic acoustic oscillations (BAO) \cite{5} have indicated that the universe is undergoing a phase of accelerated expansion, which is dubbed as dark energy (DE) by cosmologists. Recently, this mysterious phenomenon of cosmic acceleration has been confirmed, once again, by the galaxy clustering and weak gravitational lensing data from the first year (Y1) release of the Dark Energy Survey (DES) \cite{6}. In general, to characterize the properties of DE based on observational data, theoreticians propose statistically two methods, i.e., parametric and non-parametric methods. For the former case, one can specify a DE model based on some phenomenological considerations or physical mechanisms, and then constrain its parameter space using various observations. For example, up to now, the simplest DE model is the cosmological constant ($\Lambda$) plus cold dark matter (CDM) scenario, namely the so-called $\Lambda$CDM model, which has phenomenologically one typical parameter (matter density ratio $\Omega_m$) if we only consider the late-time evolution of the universe \cite{7}. What one should do is to constrain $\Lambda$CDM using data and calculate the 1-dimensional confidence interval of $\Omega_m$. The goals of constraints are: (i) for a given DE model, to determine the optimal values and statistical confidence intervals of key physical parameters; (ii) to rule out a class of DE models, which are not supported by statistical information criteria. For the latter case, without any particular parameterized form, one can reconstruct the evolution of DE or study the values of specific cosmological quantities starting directly from observations. Note that these methods must be more statistically complex than parametric ones. In the literature, there are several popular non-parametric methods such as principle component analysis (PCA) \cite{8,9}, local weighted regression (LWR) \cite{10,11} and Gaussian process (GP) \cite{12}. PCA is widely used for reducing dimension and denoising of large datasets and it has been introduced to reconstruct the equation of state (EoS) of DE \cite{13}. LWR, which is a k-nearest-neighbor-based method, is a general framework of least square regression. It has also been used for studying the dynamics of DE \cite{14}. GP is recently used with high frequencies in the fields of astrophysics and cosmology \cite{15}.
It is a stochastic process and its distribution is a distribution over functions with a continuous domain, e.g., time or space. It has been used for studying the evolution of EoS of DE \cite{16,17,18,19} and reconstruct the CMB angular power spectrum \cite{20}.

It is worth noting that previous researches all focus on testing the explanation ability (EA) of a given DE model in the usage of parametric methods, i.e., implementing the standard $\chi^2$ statistics to estimate the parameter space of typical model parameters in light of all data \cite{4}. Meanwhile, the same consequences occur in the usage of non-parametric methods, i.e., using all data, previous works all concentrate on reconstructing the underlying models with the best EA, which govern the dynamical evolution of cosmological quantities \cite{13,14,15,16,17,18,19,20}.

However, for a given set of data, the generalization ability (GA) of a model learnt from some statistical algorithm is also very important from a point of view of machine learning (ML), which is the core of artificial intelligence and data science. Usually, in practice, a ML specialist only considers the GA of a model, but one statistician just studies the EA of a model. With gradually accumulating cosmic data and improvements of computing abilities, we think that cosmologists faces a challenge, i.e., developing and using some advanced statistical method to find a DE model, which has both good EA and GA for a group of datasets. Nonetheless, whether the EA and GA of a DE model are good or not is very hard to judge and quantify simultaneously. Furthermore, the fitted DE model from data may have no reasonable physical form. To overcome naturally both difficulties, we transform the above challenge into its dual question: fixing a specific DE model such as $\Lambda$CDM, how can we find out the overlapped region of two parameter spaces from its best EA and GA in light of ML techniques ? Therefore, in this work, our task is to obtain the overlapped and reduced parameter space based on traditional $\chi^2$ statistics and ML algorithms for a given DE model and a given set of data.

\emph{Data}.--- As a performance, we adopt 30 cosmic chronometers (CC) lying in the redshift range (0, 2) as our data sample \cite{21}, which are obtained using the most massive
and passively evolving galaxies based on the `` galaxy differential age '' method, and consequently are independent of any DE model.

\emph{Methodology}.--- Firstly, we perform $\chi^2$ statistics to constrain $\Lambda$CDM using CC, and let the obtained parameter space via the Markov Chain Monte Carlo (MCMC) technique correspond to the best EA of $\Lambda$CDM. Secondly, through separating the whole sample into training set and testing set, we take seven different ML algorithms to learn CC data, and use the performance measure, mean squared error (MSE), to choose the optimal ML algorithm, which will be used for obtaining the parameter space corresponding to the best GA of $\Lambda$CDM via MCMC method. Furthermore, we can find out the overlapped parameter space for $\Lambda$CDM. Then, we shall repeat this process for non-flat $\Lambda$CDM ($o\Lambda$CDM) and $\omega$CDM models, where $\omega$ is constant EoS of DE.

Specifically, 7 ML algorithms are, respectively, ordinary least square (OLS), least absolute shrinkage and selection operator (LASSO) \cite{22}, ridge regression (RR) \cite{23}, elastic net (EN) \cite{24}, linear support vector regression (LSVR) \cite{25}, neural network (NN) \cite{26} and Bayesian ridge regression (BRR) \cite{27}. To choose the optimal ML algorithm, at first, except for OLS, we select a specific value for the hyperparameter according to our previous experiences, and take randomly 3-fold cross validation to train data and test the obtained model for each algorithm. Subsequently, minimizing the loss function (LF) on training set, we can get a model with best-fitting values of parameters, which shall be used for calculating MSE on testing set. Then, we implement the process of hyperparameter tuning and choose the best value of hyperparameter, which makes MSE on testing set minimal. Repeating this program 3 times and averaging 3 MSEs, we can have the best MSE, which is regarded as our score for each algorithm. Finally, we select the optimal algorithm which has the least score among 7 algorithms.

When applying ML to CC, we first build a bridge between ML and $\chi^2$ statistics by identifying the LF in ML with the likelihood function in $\chi^2$. Since one in the field of ML often deals with data without errors but cosmological data such as CC has errors, we then add errors of CC, for the first time, into the LF by fitting CC with a power-law polynomial (PLP), where we add errors of Hubble parameters $H(z)$ at each redshift $z$ into each term of PLP. Note that, to extract utmostly the information hiding in CC, we have taken a feature adding technique via a PLP to express $H(z)$, since there is only 1 feature for CC using ML terminology. Subsequently, we propose a conjecture:

`` \emph{The model learnt from data is equivalent to the simplest one constructed from physical considerations} ''.

For the case of CC, it can be shown as `` $H_{\mathrm{ML}}(z)= H_{\Lambda\mathrm{CDM}}(z)$ ''  (or $H_{o\mathrm{\Lambda CDM}}(z)$, $H_{\mathrm{\omega CDM}}(z)$):
\begin{equation}
\omega_0+\omega_1z+...+\omega_{n_d}z^{n_d}=H_0\sqrt{\Omega_m(1+z)^3+1-\Omega_m},
\label{4}
\end{equation}
where $H_0$, $n_d$ and $\omega_i$ ($i\in$ [0, $n_d$]) denote the Hubble constant, the number of degree and the coefficients of PLP, respectively. To determine the 2-dimensional parameter space ($H_0$ versus $\Omega_m$ ) of $\Lambda$CDM, we must first determine $n_d$ of $H_{\mathrm{ML}}(z)$, which equals the number of terms $n_t$ of $H_{\mathrm{ML}}(z)$ minus 1. Then, we Taylor expand the above equation on both sides around $z=0$ and can naturally reexpress the coefficients $\omega_i$ of PLP with $H_0$ and $\Omega_m$. Furthermore, inserting the reexpressed $\omega_i$ into the LF with optimal hyperparameters, we minimize the LF on the whole sample and thereby obtain the parameter space from ML, which represents the best GA of $\Lambda$CDM.
( The basic formula used in this work and all the mathematical details of 7 ML algorithms can be found in the supplementary materials.)

\begin{figure}
\centering
\includegraphics[scale=0.4]{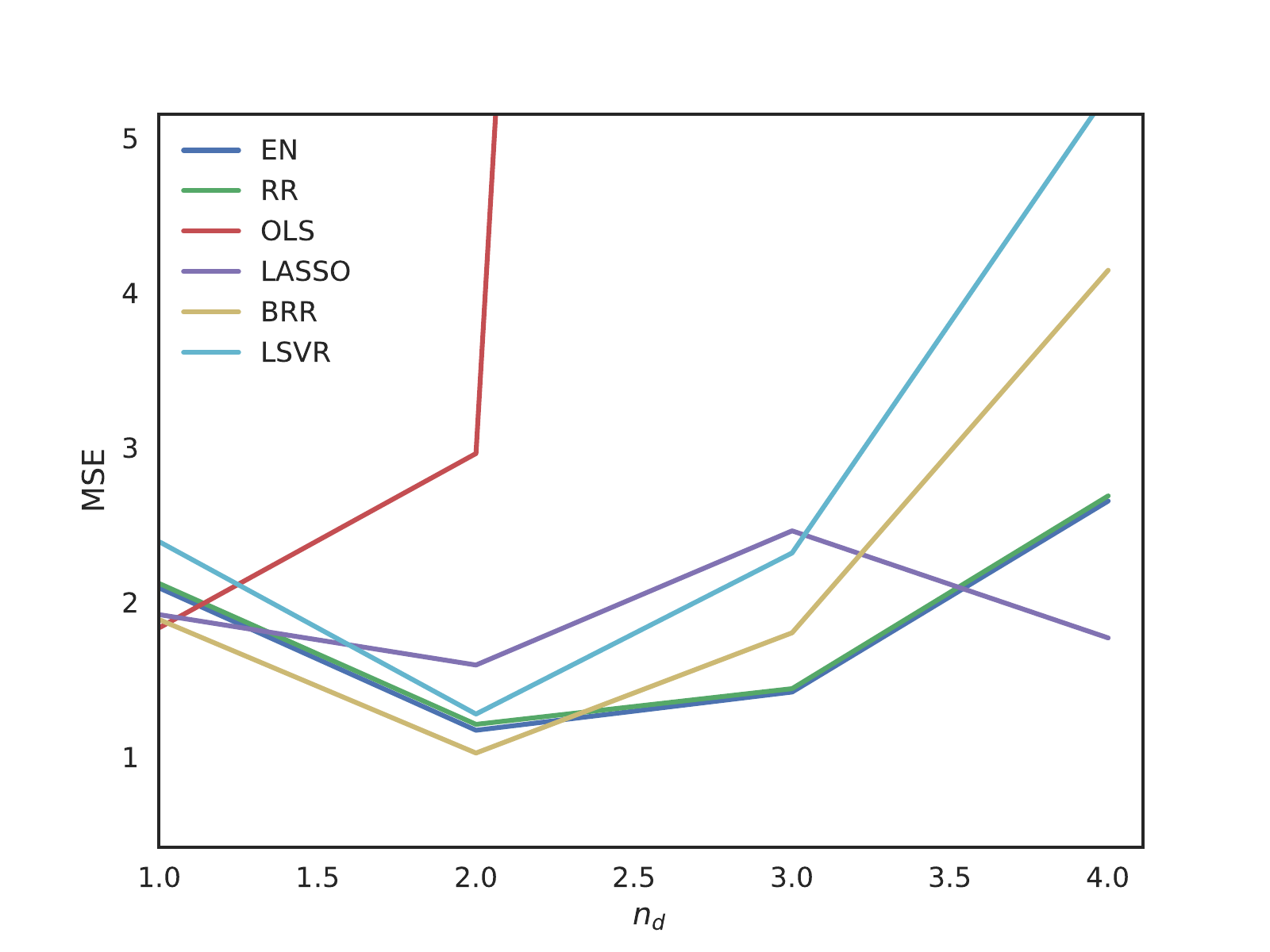}
\includegraphics[scale=0.4]{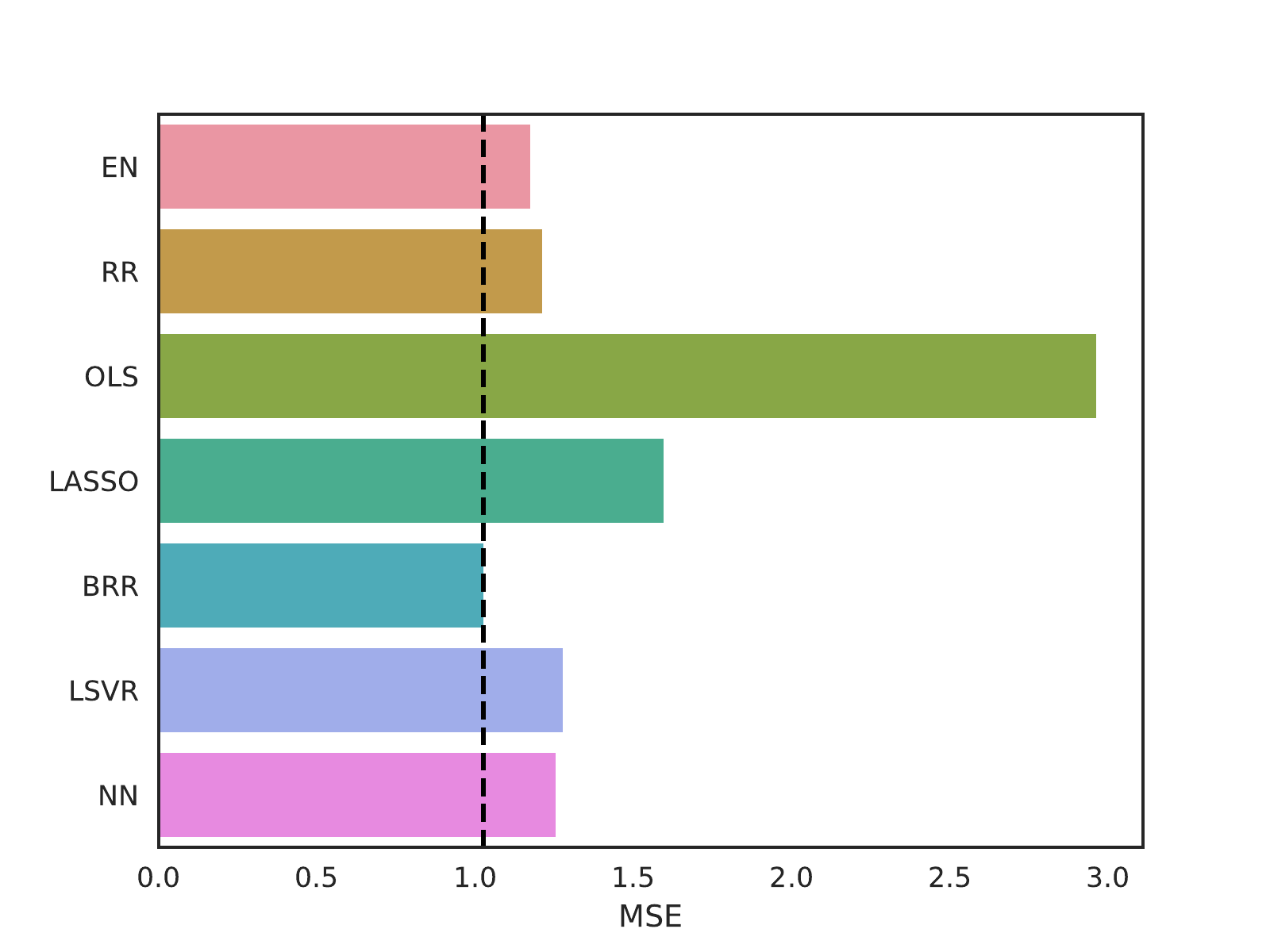}
\caption{\emph{The upper panel}: relationships between MSE on testing set and $n_d$ of each ML algorithm except for NN. \emph{The lower panel}: on testing set, values of MSEs of 6 ML algorithms at the best degree $n_d=2$ and that of NN after hyperparameters being tuned to the optimal are shown, and the black (dashed) line is the value of MSE of BRR.}\label{f1}
\end{figure}

\begin{figure}
\centering
\includegraphics[scale=0.4]{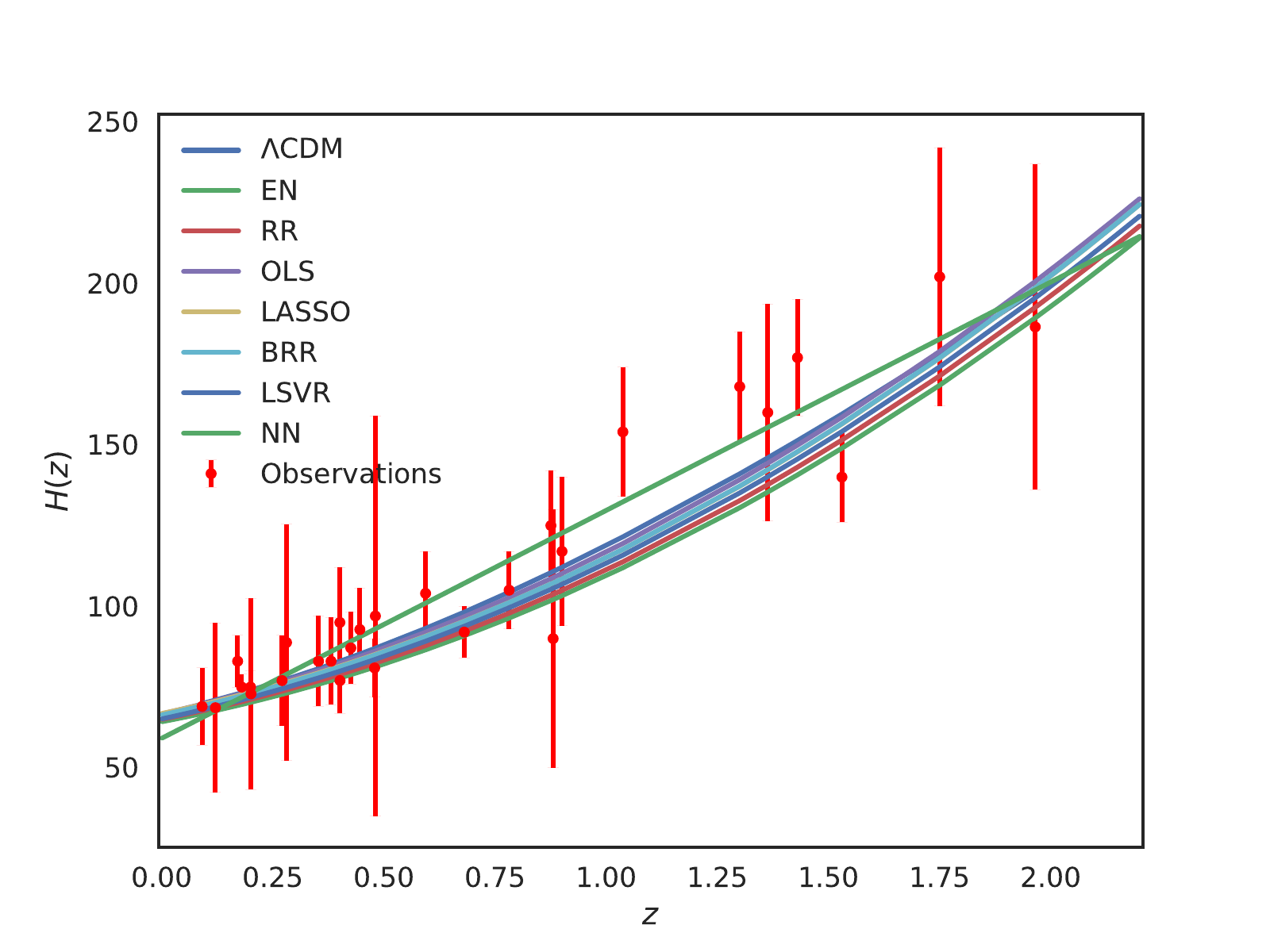}
\caption{The comparison between the best-fitting models from 7 ML algorithms using CC data and the best-fitting $\Lambda$CDM model from Planck CMB data \cite{4,28}. The points with red error bars are CC data.}\label{f2}
\end{figure}

\begin{figure}
\centering
\includegraphics[scale=0.4]{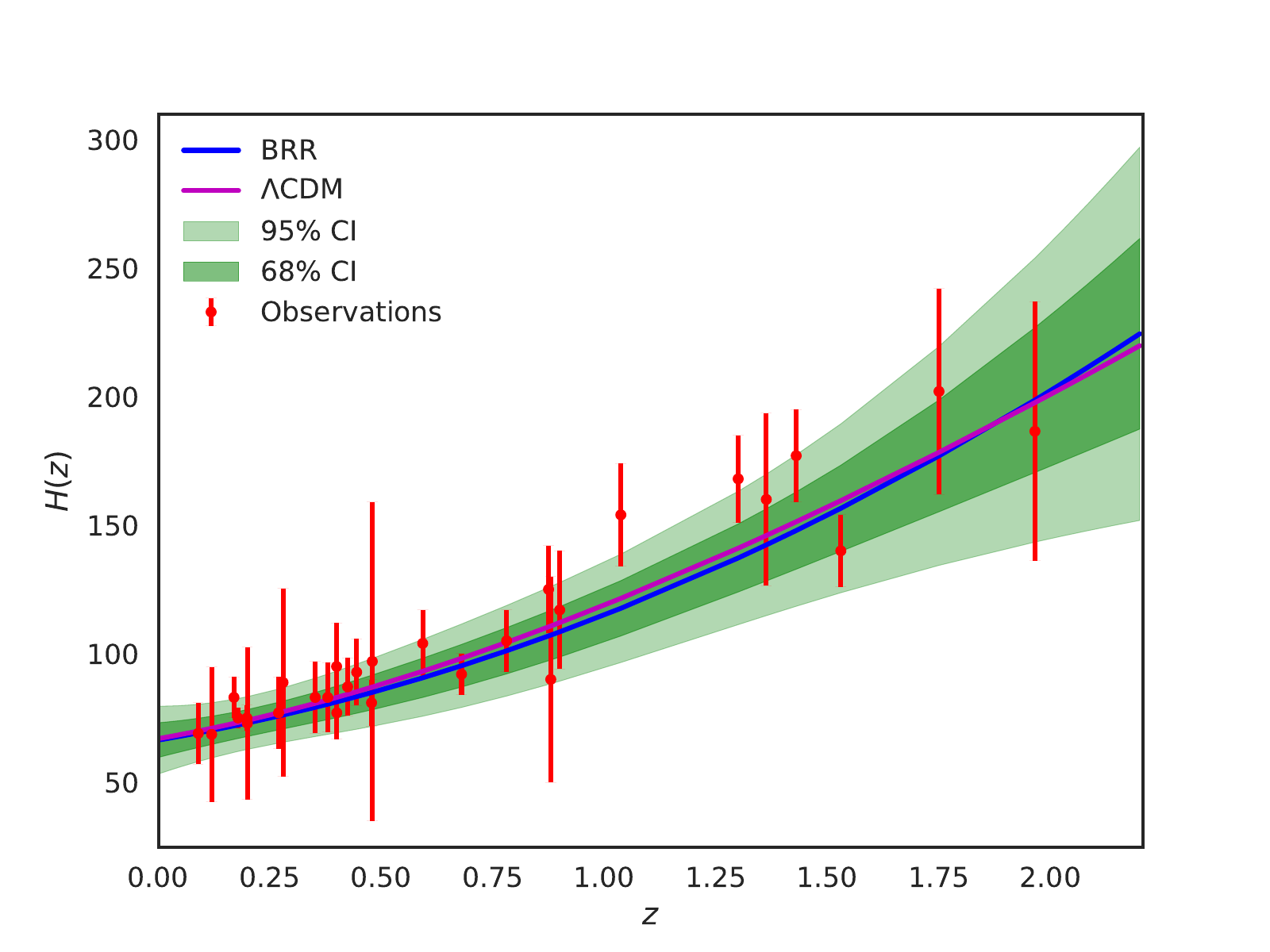}
\caption{The comparison between the best-fitting model, 1$\sigma$, 2$\sigma$ confidence bands from BRR using CC data and the best-fitting $\Lambda$CDM model from Planck CMB data \cite{4,28}.}\label{f3}
\end{figure}

\begin{figure*}
\centering
\includegraphics[scale=0.36]{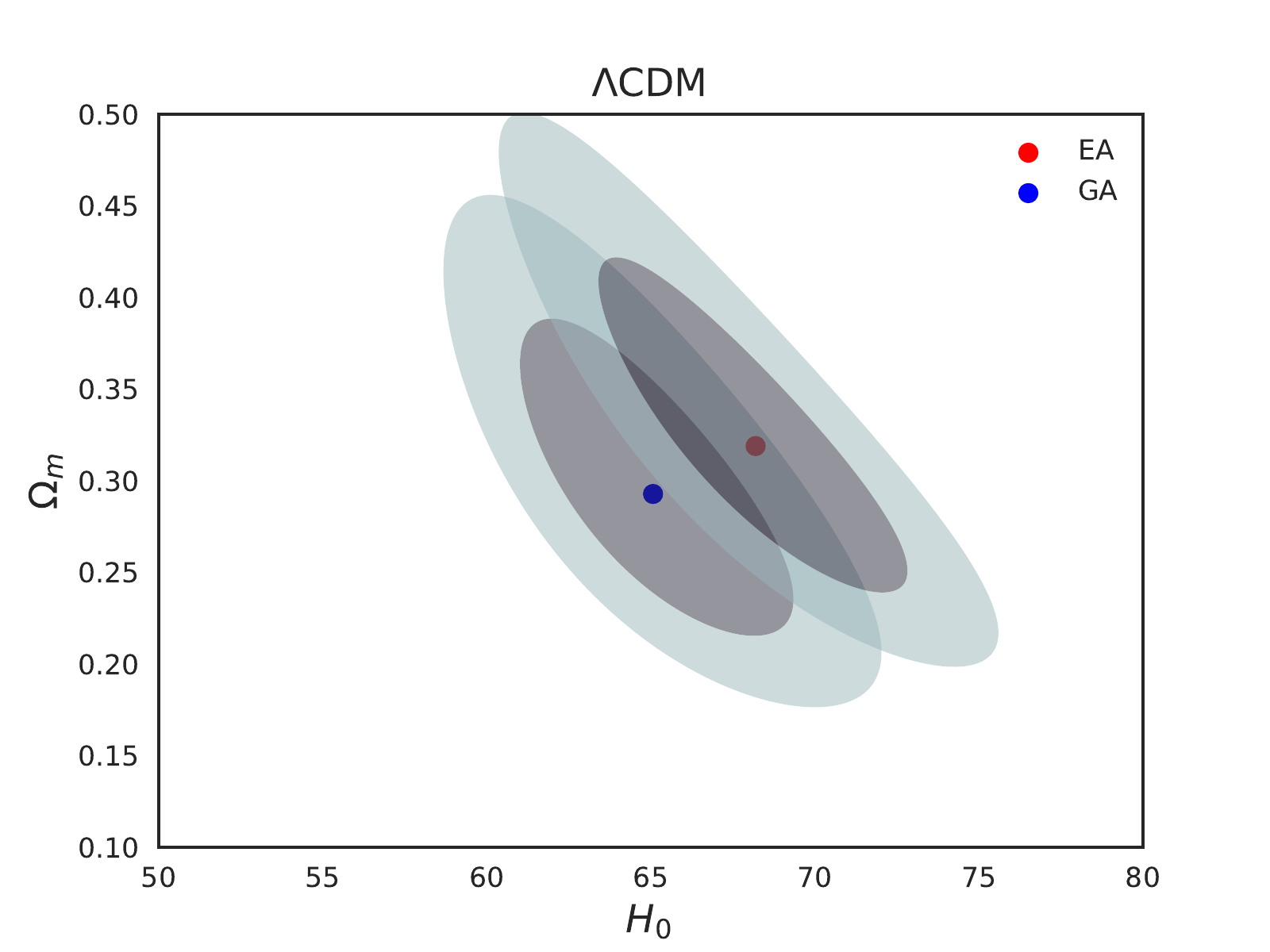}
\includegraphics[scale=0.36]{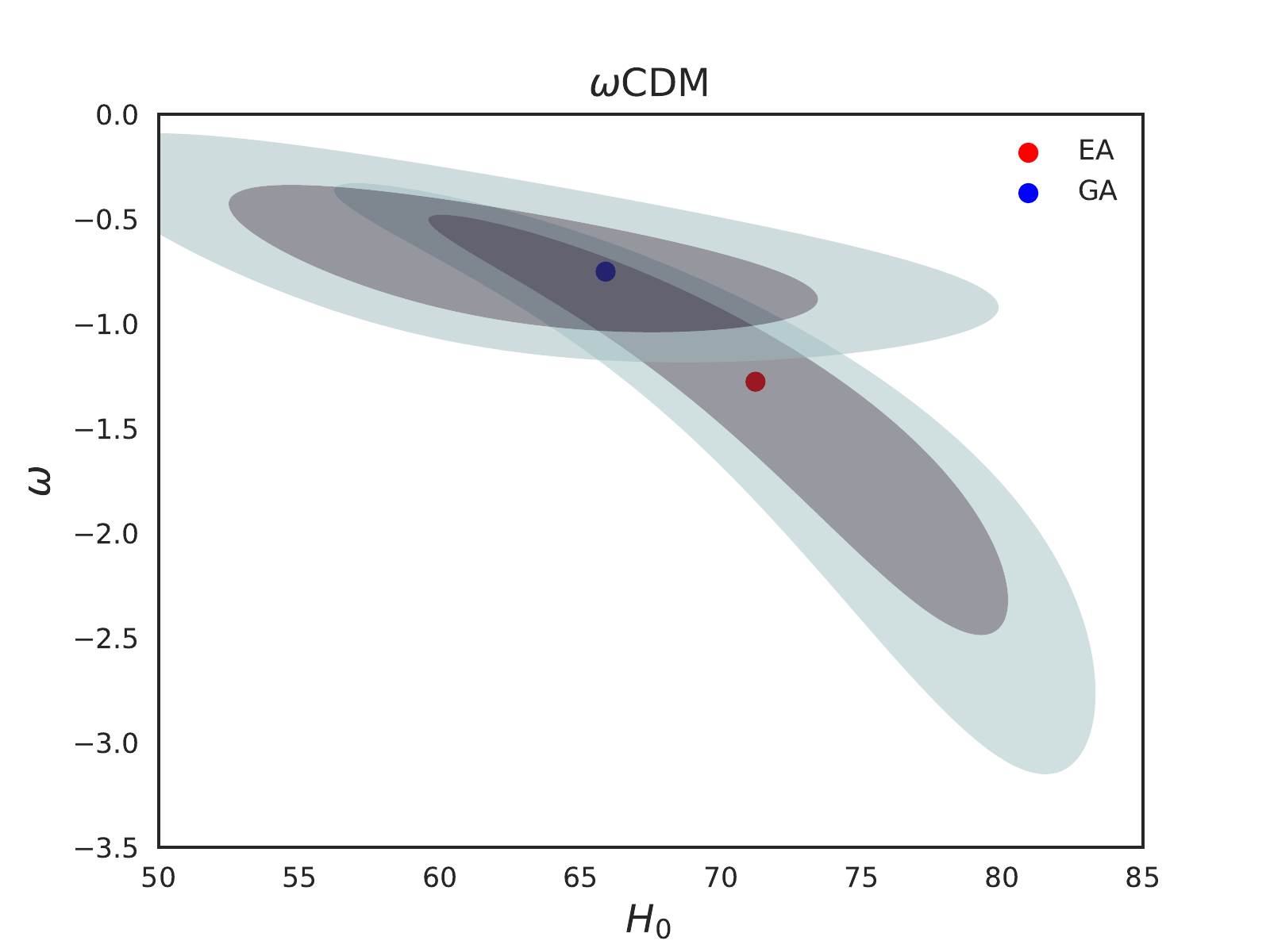}
\includegraphics[scale=0.36]{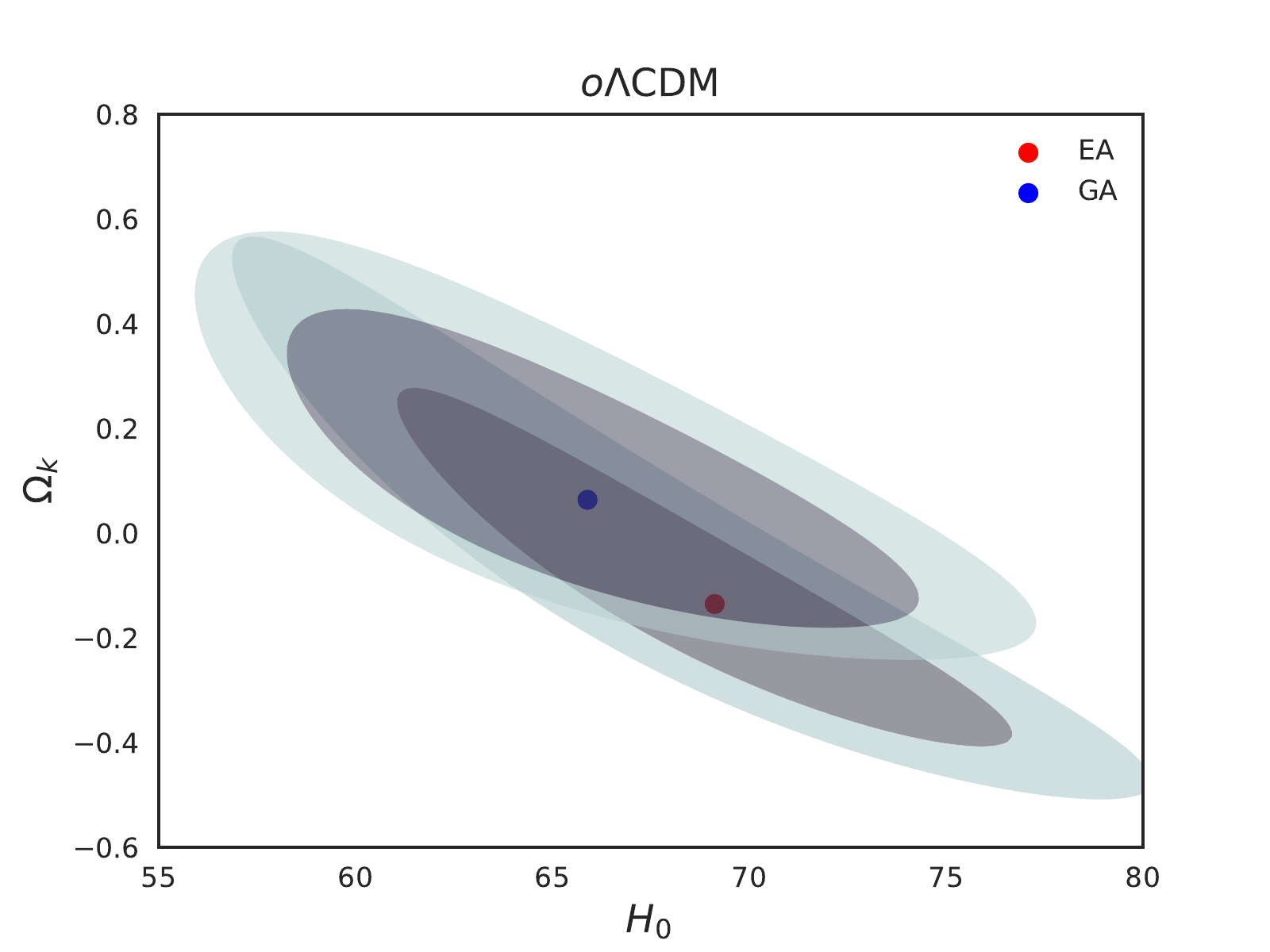}
\caption{From left to right, in light of CC data, the EPSs and GPSs of $\Lambda$CDM, $\omega$CDM and $o\Lambda$CDM models are presented, respectively. The red and blue points are best-fitting ones corresponding to EAs and GAs, respectively.}\label{f4}
\end{figure*}

\emph{Results}.--- For 6 parametric ML algorithms, using CC, we find that the best degree $n_d=2$ for a wide range $n_d\in$ [1, 30] and this result is very stable for 3-fold data separation, and that the zeroth-order term of each PLP is $H_0$. Meanwhile, in the upper panel of Fig. \ref{f1}, the optimal algorithm which has the least MSE 1.022 among 6 algorithms is BRR. Since we take conventionally OLS to confront cosmological models with observations, OLS should be used as as an important reference. We find that OLS has the largest MSE 2.957 and performs the worst at $n_d=2$. It is noteworthy that MSE of LASSO obeys a different variation tendency in comparison with those of 5 left methods, i.e., it suddenly decreases at $n_d=4$. Interestingly, its value at $n_d=2$ is still less than that at $n_d=4$ by a difference 0.162. The MSE of LSVR 1.291 is a little larger than those of EN and RR, which are very similar at all degrees.

For completeness and a naive comparison to 6 parametric methods, we also introduce the non-parametric error-back-propagation NN algorithm, which consists of 1 input layer (1 neuron), 2 hidden layers (30 neurons per layer) and 1 output layer (1 neuron). Note that for the first time in cosmology, we have successfully adding data errors into NN, which is very important to maximize the information utilization. After careful hyperparameter tuning, we find that BRR is still the optimal algorithm (see the lower panel of Fig. \ref{f1}). Hence, using CC, we choose BRR as our best ML method to derive the generalization parameter space (GPS) of $\Lambda$CDM.

The relationships between $H(z)$ and $z$ from 7 ML algorithms only using CC and $\Lambda$CDM (here we use $H_0=66.93$ km s$^{-1}$ Mpc$^{-1}$ \cite{28} and $\Omega_m=0.308$ \cite{4}) constrained by Planck CMB angular power spectrum data are shown in Fig. {\ref{f2}}. We find that the best-fitting models of 6 parametric algorithms are very close to each other as well as $\Lambda$CDM and that of our non-parametric NN exhibits an interesting straight-line-like behavior. Furthermore, in Fig. \ref{f3}, we present 1$\sigma$ ($\%68$) and 2$\sigma$ ($\%95$) confidence bands from our best BRR method, and obtain the first Hubble constant estimation $H_0=65.95^{+6.98}_{-6.36}$ km s$^{-1}$ Mpc$^{-1}$ from BRR via a $n_d=2$ PLP by only considering the GA of the model learnt from data. Although this new estimation is consistent with the indirectly global derivation by Planck Collaboration \cite{28} and directly local measurement by Riess \emph{et. al.} \cite{29} at the 1$\sigma$ confidence level (CL), its optimal value prefers the Planck's derivation. Note that this estimation is not only very compatible with our previous result $H_0=67.38\pm4.72$ km s$^{-1}$ Mpc$^{-1}$ based on GP at the 1$\sigma$ CL by using the same CC data \cite{30}, but also has larger uncertainties than it does. In addition, one can also easily find that the best-fitting $\Lambda$CDM from Planck data is very consistent with the underlying model from BRR at the 1$\sigma$ CL.

As described above, optimizing the reexpressed LF from BRR, we obtain, respectively, the best-fitting values and 1$\sigma$ uncertainties of parameters corresponding to GAs of $\Lambda$CDM, $\omega$CDM and $o\Lambda$CDM models. Specifically, we have $H_0=65.07^{+4.35}_{-4.06}$ km s$^{-1}$ Mpc$^{-1}$
, $\Omega_m=0.292^{+0.094}_{-0.080}$ for $\Lambda$CDM, $H_0=65.89^{+7.68}_{-13.35}$ km s$^{-1}$ Mpc$^{-1}$, $\omega=-0.751^{+1.096}_{-0.304}$ for $\omega$CDM, and $H_0=65.89^{+8.47}_{-7.61}$ km s$^{-1}$ Mpc$^{-1}$, $\Omega_k=0.064^{+0.363}_{-0.247}$ for $o\Lambda$CDM. Subsequently, we find that: (i) although the values of $H_0$ from GAs of 3 models are consistent with those from their EAs at the 1$\sigma$ CL, the best-fitting $H_0$ values from GAs are all less than those from EAs; (ii) the values of $\Omega_m$, $\omega$ and $\Omega_k$ for 3 models are, respectively, compatible with Planck's results at the 1$\sigma$ CL \cite{4}; (iii) for $\omega$CDM, considering GA, $\omega$ is weakly anti-correlated with $H_0$, which is different the case of EA; (iv) all the values of parameters have large errors due mainly to small data sample size.

Plotting the explanation parameter space (EPS) from $\chi^2$ statistics together with the GPS from BRR, we obtain the overlapped and reduced parameter spaces of $\Lambda$CDM, $\omega$CDM and $o\Lambda$CDM models, respectively (see Fig. \ref{f4}). Furthermore, we specify these narrow and overlapped regions at the 2$\sigma$ CL as follows: $H_0\in$ [60.52, 72.14], $\Omega_m\in$ [0.206, 0.454] for $\Lambda$CDM, $H_0\in$ [56.28, 74.49], $\omega\in$ [-1.206, -0.335] for $\omega$CDM, $H_0\in$ [56.86, 75.97], $\Omega_k\in$ [-0.255, 0.560] for $o\Lambda$CDM. From the view of statistics, we find that $o\Lambda$CDM from EA has a stronger GA than $\Lambda$CDM and $\omega$CDM from EAs, since the area ratio $R$ between the overlapped contour and the EA contour of $o\Lambda$CDM is larger than $50\%$ ($R<50\%$ for both $\Lambda$CDM and $\omega$CDM). From Fig. \ref{f4}, one can also find that the 1$\sigma$ interval of $H_0$ of overlapped parameter space of $\Lambda$CDM is smaller than those of $\omega$CDM and $o\Lambda$CDM. This indicates that the situation that the EPS of $\Lambda$CDM should become larger than before by adding a physical parameter $\omega$ or $\Omega_k$ to it will also occur to its overlapped parameter space. In total, making full use of CC, we have verified that the parameter spaces of 3 cosmological models can be obviously compressed (or reduced) by considering both their EAs and GAs.

\emph{Discussions}.--- Using CC, our best degree $n_d=2$ for 6 algorithms may imply that, from the view of ML, we have ruled out a large class of models where the number of physical parameters $n_p>3$ including $H_0$. This is why we just choose the $\Lambda$CDM ($n_p=2$), $\omega$CDM ($n_p=3$) and $o\Lambda$CDM ($n_p=3$) models to derive the overlapped parameter spaces. Based on the consideration that the ultimate aim using data to constrain DE models is to reduce ceaselessly their EPSs, discard all the impossible models and give the more accurate values of parameters of correct models than before, to a large extent, we can reduce the current EPSs if we also consider the GPSs of DE models. From Fig. \ref{f1}, we know that BRR in all 7 ML algorithms is the best algorithm to characterize the GA of 3 different cosmological models for CC data. But if one is dealing with other datasets such as SNIa, LSS and CMB, the best ML method may not be BRR. Although overlapped parameter spaces of the above 3 DE models are all found out by using CC, we cannot ensure that the EPS and GPS of any given DE model with $1<n_p\leqslant3$ have a common region.

\emph{Conclusions}.--- We have successfully introduced the ML techniques into the field of cosmology by considering simultaneously the EA and GA of a specific DE model. For the first time, we add the information of data errors into the usage of 7 ML algorithms via a PLP expansion when studying the background evolution of the universe.
We make a connection between ML and traditional $\chi^2$ statistics by use of a physical example, i.e., using BRR to find out the overlapped regions of EPSs and GPSs of $\Lambda$CDM, $\omega$CDM and $o\Lambda$CDM models, respectively. For the first time, using CC, we find that the parameter spaces of 3 DE models can be clearly reduced by considering both their EAs and GAs. Very interestingly, from the view of statistics, we find that $o\Lambda$CDM has a stronger GA than $\Lambda$CDM and $\omega$CDM by performing $\chi^2$ fitting to CC data.

\emph{Prospects}.--- With fast growing astronomical data, continuously improved data-storage capabilities and computing abilities, there is no doubt that ML will play a very important role in the researches of astronomy, astrophysics and cosmology. We expect that ML methods can extract new and useful information about underlying physical laws from future huge amounts of data.

\emph{Acknowledgements}.--- We thank Jing-Ling Chen, Wu-Sheng Dai and Yu-Xiao Liu for useful communications.

\end{document}